\documentclass[aps,prc,twocolumn,showpacs,superscriptaddress,groupedaddress]{revtex4-1}  % for review and submission
\usepackage{mathrsfs,amsmath,amssymb,amsthm}
\usepackage{amssymb,fmtcount}
\usepackage{graphicx,epsfig,latexsym,overpic,amssymb,color}
\usepackage{threeparttable}
\usepackage{lineno,hyperref}
\preprint{\today}

\newcommand{\rr} {\boldsymbol{r}}

\begin{document}
\title{
Correlations between nuclear landscape boundaries and neutron-rich $r$-process abundances
}
\author{Q.Z. Chai}
\affiliation{
State Key Laboratory of Nuclear Physics and Technology, School of Physics,
Peking University, Beijing 100871, China
}
\author{Y. Qiang}
\affiliation{
State Key Laboratory of Nuclear Physics and Technology, School of Physics,
Peking University, Beijing 100871, China
}
\author{J.C. Pei}\email{peij@pku.edu.cn}
\affiliation{
State Key Laboratory of Nuclear Physics and Technology, School of Physics,
Peking University, Beijing 100871, China
}

%\author{D.W. Guan}
%\affiliation{
%State Key Laboratory of Nuclear Physics and Technology, School of Physics,
%Peking University, Beijing 100871, China
%}

%\author{Min-Liang Liu}
%\affiliation{
%Institute of Modern Physics, Chinese Academy of Sciences, Lanzhou 730000,  China
%}
%\author{Fu-Rong Xu}
%\affiliation{
%School of Physics, Peking University, Beijing 100871,                      China
%}
%\affiliation{
%Institute of Theoretical Physics, Chinese Academy of Sciences, Beijing   100080,
%                                                                           China
%}
%\affiliation{Center of Theoretical Nuclear Physics, National Laboratory of Heavy
%                                           Ion Collisions, Lanzhou 730000, China
%}
%123456789 123456789 123456789 123456789 123456789 123456789 123456789 123456789
\begin{abstract}
Motivated by the newly observed $^{39}$Na in experiments,  systematic calculations of global nuclear binding energies with seven Skyrme forces
are performed. We demonstrate the strong correlation between the two-neutron separation energies ($S_{2n}$) of $^{39}$Na and the total number of bound nuclei of the whole
nuclear landscape. Furthermore, with calculated nuclear masses, we perform astrophysical rapid-neutron capture process ($r$-process) simulations  by using nuclear
reaction code TALYS and nuclear reaction network code SkyNet. $r$-process abundances from ejecta of neutron star mergers
and core-collapse supernova are compared.
Prominent covariance correlations between nuclear landscape boundaries and  neutron-rich $r$-process abundances before the third peak are shown.
This study highlights the needs for further experimental studies of drip-line nuclei around $^{39}$Na   for better constraints on nuclear landscape boundaries and $r$-process.

%1234567890123456789012345678901234567890123456789012345678901234567890123456789
\end{abstract}
\pacs{  21.10.Re, 21.60.Cs, 21.60.Ev}
\maketitle
%123456789 123456789 123456789 123456789 123456789 123456789 123456789 123456789
\section{introduction}

It is well known that studies of exotic nuclei close to drip lines are
precious for understandings of the origin of elements
in nature~\cite{Cowan2021,Martin2016}.
The rapid neutron capture process ($r$-process)
involving high neutron densities is responsible for producing
half of the elements heavier than iron and all elements beyond bismuth.
The actual astrophysical sites for the occurrence of $r$-process is not definitely determined
yet~\cite{Cowan2021,Arnould2007,Kajino2019}.
The developments of new-generation rare isotope beam facilities around the world provide
unprecedent opportunities to access the nuclear drip lines.
For example, the Facility for Rare Isotope Beams (FRIB) is expected to be
fully operational in 2022 and will be able to reach the neutron
drip line up to nuclei with charge number $Z$ = 40~\cite{fribc}.
However, it is almost impossible to reach the neutron drip line in heavy nuclear mass region by
terrestrial experiments. Therefore, the examination on correlations between existing experimental evidences
and theoretical predictions is crucial for better exportations.

In a very recent experiment performed in RIKEN, $^{31}$F and $^{34}$Ne were reconfirmed to be the
drip-line nuclei~\cite{Ahn2019PRL}.
Surprisingly, this experiment also observed one event of $^{39}$Na~\cite{Ahn2019PRL},
indicting it is weakly bound and mostly
likely it is the drip line of sodium.
This is an exciting progress to reach the neutron drip line since the last observation of $^{40}$Mg
in 2007~\cite{Baumann2007}.
It is known that different theoretical models can have
remarkably divergent predictions about the neutron
drip lines~\cite{Erler2012,WangNing2014,ChenLW2015,Goriely2016}.
In contrast, the proton drip line has much reduced uncertainties.
Therefore, the newly observed $^{39}$Na provides a great opportunity to constrain theoretical models.
It is interesting to know how small discrepancies in drip lines of light nuclei propagate to
large uncertainties in drip lines of heavy nuclei.
Consequently, the total number of bound nuclei in the nuclear landscape
can be more accurately estimated.

So far it was known that binary neutron star mergers (NSM)~\cite{Freiburghaus1999,Korobkin2012}
and ejecta from core collapse supernova (CCSN)~\cite{Qian2000,Winteler2012,Mosta2018}
are possible scenarios for $r$-process.
Following the gravitational wave event GW170817 of NSM,
the $r$-process kilonova electromagnetic transient was observed, resulting from the ejection of $\thicksim$0.05
solar masses of neutron-rich material~\cite{Drout2017}.
These observations are becoming increasingly precise.
NSM provides a much higher neutron density scenario to support a strong version of
$r$-process to reach heavy elements such as uranium and thorium,
 while CCSN is associated with a larger electron faction $Y_e$.
Therefore it is expected that $r$-process via NSM is more sensitive to
properties of neutron drip lines compared to that via ejecta of CCSN.
There have been extensive studies of the impacts of uncertainties of nuclear inputs for $r$-process abundances
in the literature~\cite{Martin2016,Mumpower2016,SBH2008,NZM2019,Sprouse2020}.
The $r$-process involves the neutron capture reactions, $\beta$-decays and fission properties.
The fission is essential for appearance of the second peak ($A$$\sim$160) in elemental abundance~\cite{Eichler2015,Goriely2013}.
The uncertainties in (n, $\gamma$) reactions play a sensitive role.
It was found that mass
variations of $\pm$0.5 MeV can result in up to an order of magnitude change in the final abundance~\cite{Mumpower2016}.
The (n, $\gamma$) reaction is mainly determined by the nuclear masses and thus
reliable predictions of nuclear mass by self-consistent microscopic framework are crucial.

In principle, the $ab$ initio calculations of nuclear drip lines are  more reliable
but it is problematic for heavy nuclei due to tremendous computing costs~\cite{Holt2021}.
Semi-microscopic and phenomenological models can be precise for known nuclei but could be less reliable
for explorations.
As a suitable tool, the density functional theory with high-precision effective interactions
are versatile for reasonable descriptions of global finite nuclei and neutron stars,
including exotic structures and dynamics of halo nuclei,  and
nuclear fission~\cite{Erler2012,Bender2003,jcj2021,qy2021,zzw2018,wang2017,xxy2016,zy2014,Chai2020}.
Previously the properties of $^{39}$Na and neighboring drip line nuclei have been studied~\cite{Chai2020}.
The subsequent combined constrains on the whole nuclear landscape and $r$-process are expected.

Compared to earlier studies of $r$-process by focusing on the impact of uncertainties of
nuclear inputs~\cite{Mumpower2016},
the aim of this work is to examine the correlations between theoretical discrepancies in $^{39}$Na,
nuclear landscape boundaries, and $r$-process abundances based on several
effective nuclear forces.
Firstly, the global nuclear masses are calculated with the Skyrme Hartree-Fock-Bogoliubov (HFB)
framework~\cite{Perez2017}.
In particular, the results are evaluated with the existing evidence of the drip line nucleus $^{39}$Na.
This results in very different total number of bound nuclei and $r$-process paths.
With the calculated nuclear masses, the (n, $\gamma$) reaction rates are obtained with
TALYS~\cite{talys1.95}.
The updated reaction rates are merged into REACLIB database~\cite{Cyburt2010}, and then
the $r$-process simulations are performed with SkyNet~\cite{Jonas2017}
which interfaces with REACLIB.
Finally, the covariance correlations between $r$-process abundances
and nuclear landscape boundaries are analyzed.

\section{The theoretical framework}
%%%%%%%%%%%%%%%%%%%%%%%%%%%%%%%%%%%%%%%%%%%%%%%%%%%%%%%%%%%%%%%%%%%%%%%%%%%%%%%%
%123456789 123456789 123456789 123456789 123456789 123456789 123456789 123456789

Firstly, the global nuclear masses at ground states are calculated by
the Skyrme HFB approach with the parallel scheme.
The HFB calculations are performed with the HFBTHOv3.00 solver~\cite{Perez2017},
in which wavefunctions are presented by the basis expansion of 22
harmonic oscillator shells.
The default oscillator length $b_0 = \sqrt{\hbar/m\omega_0}$, where
$\hbar \omega_0 = 1.2 \times 41/A^{1/3}$.
For each nucleus, the ground state is determined by computing several
quadrupole deformations from $\beta_2$=$-$0.5 to 0.5,
in case shape coexistence present.

In HFB calculations, seven Skyrme type effective forces have
been adopted.
SkM$^*$ force has good surface properties
and has been widely applied in descriptions of fission~\cite{Bartel1982}.
SLy4 force has been widely used in descriptions of neutron-rich nuclei and
neutron stars~\cite{Chabanat1998}.
SLy4$'$ force is a refitted force that improves global descriptions of binding energies
compared to the original SLy4~\cite{xxy2016}.
UNEDF0 has been well optimized for descriptions of global binding energies
with a high precision~\cite{unedf0}.
In addition, we speculate that a single density dependent term in standard Skyrme forces
is not sufficient for the Skyrme force to simulate  many-body correlations.
The extended SLy4E$_{\rm global}$~\cite{xxy2016},
SkM$^*$$_{\rm ext1}$ and UNEDF0$_{\rm ext1}$ forces~\cite{zzw2018}
 with an additional high-order density-dependent term are also adopted.
In the particle-particle channel, a density-dependent pairing interaction
has been adopted~\cite{wang2017,pair2008},
\begin{equation}
    V_{pair}(\rr)
    =
    V_0
    \left\{1-\eta {\left[\frac{\rho(r)}{\rho_0}\right]}^\gamma \right\},
                                                                  \label{eqn03}
\end{equation}
where $\rho_0$ is the nuclear saturation density and
we adopt the constants as $\eta=0.8$ and $\gamma=0.7$.
The pairing strengths $V_0$ are fitted to the neutron
gap of $^{120}$Sn of 1.245 MeV for different Skyrme forces.
The pairing gaps could be very different towards drip lines by using different pairing interaction forms.
The resulted pairing gaps  at the neutron drip lines are between the surface pairing and
the mixed pairing~\cite{wang2017}.
This is a reasonable choice because the pairing gaps  obtained  with the surface pairing interaction
are too large toward the neutron drip lines if the pairing strength is invariant for stable
and weakly bound nuclei.
The global binding energies of odd-$A$ and odd-odd nuclei are obtained by
the average pairing gap method after even-even nuclei are calculated with
the HFB approach~\cite{Erler2012,ChenLW2015,Bayesian2020}.

Secondly, we compute neutron capture rates with the TALYS code~\cite{talys1.95}
 and the calculated nuclear masses.
The neutron capture rate is sensitive to the neutron separation energy~\cite{Mumpower2016}.
Calculated masses are used in TALYS when no experimental masses are available.
The reaction rates are calculated at 24 temperatures ranging from
$T_9$ = 0.1 to 10 GK.
The reaction rates $\lambda$ are converted to coefficients $a_0\sim a_6$ in REACLIB format~\cite{Cyburt2010},
\begin{equation}
\lambda={\rm exp}(a_0+a_1 T_9^{-1} +a_2 T_9^{-1/3} + a_3 T_9^{1/3}+a_4 T_9
+a_5 T_9^{5/3} +a_6 {\rm In} T_9 ),
\end{equation}
where $a_0\sim a_6$ are obtained by the least square fitting method and
next we updates the REACLIB data.
The inverse ($\gamma$, n) reaction rates are calculated with the detailed balance~\cite{Jonas2017}.
In this work, we replaced 3825 (n, $\gamma$) reaction rates for targets with $10 \leq Z \leq 83$
and 2453 (n, $\gamma$) reaction rates for targets with $84 \leq Z \leq 112$
in REACLIB.
It was reported that $r$-process abundances are less sensitive to uncertainties of $\beta$-decay rates
compared to neutron capture rates~\cite{NZM2019}.
The present r-process nucleosynthesis calculation includes
7836 nuclear species and 95051 reactions rates.
In SkyNet, the nuclear statistical equilibrium (NSE) is adopted for all strong reactions when
$T_9 \geq 7.0$ GK~\cite{Jonas2017}.
The NSE is calculated with a given temperature, density and $Y_e$ in SkyNet.
This is different from WinNet and XNet in which inverse rates taken from REACLIB are not completely
consistent with NSE~\cite{Jonas2017}.

Finally, the abundance evolution is calculated with SkyNet,
which actually solves the reaction network equations, i.e.,  the coupled
first-order non-linear ordinary differential equations, with a given set of rates~\cite{Jonas2017}.
The initial NSE abundances are obtained with given temperature $T$, entropy $S$ and $Y_e$.
The initial density $\rho$ is related to entropy that is proportional to $T^3$/$\rho$.
After the numerical convergence is obtained at
the evolution time of 10$^9$ s ($T$$\approx$3$\times$10$^{5}$ K, $Y_e$$\approx$0.4),
the final abundance are obtained by the sum over all reaction species.
In this work, for the ejecta of NSM, the initial temperature is taken as 7.1 GK;
$Y_e$ is taken as 0.03 (within ranges suggested in ~\cite{Korobkin2012});
and initial density is taken as 2.2$\times$10$^{11}$ g cm$^{-3}$ ($S$=2.8 $k_{\rm B}/$baryon).
For the ejecta of CCSN, the initial temperature is taken as 10 GK;
$Y_e$ is taken as 0.2 according to~\cite{Mosta2018,Lippuner2015};
and initial density is taken as 2.0$\times$10$^{8}$ g cm$^{-3}$ ($S$=10 $k_{\rm B}/$baryon).
The density expansion timescales of the ejecta are taken as 1 ms and 20 ms for NSM and CCSN respectively.
The combination of very low $Y_e$ and rapid expansion timescale guarantees the occurrence of
a strong $r$-process~\cite{Cowan2021}.
It is difficult to reproduce the solar $r$-process abundances by only one $r$-process scenario.
 SkyNet is a flexible modular library and  has been successfully used for nucleosynthesis calculations in all astrophysical scenarios~\cite{Jonas2017}.
For example, very recently, Jin \emph{et al.} have investigated that the enhanced triple-$\alpha$
reaction reduces proton-rich nucleosynthesis in supernovae using SkyNet~\cite{JinSL2020}.

\section{Results and discussions}
%%%%%%%%%%%%%%%%%%%%%%%%%%%%%%%%%%%%%%%%%%%%%%%%%%%%%%%%%%%%%%%%%%%%%%%%%%%%%%%%
%%%%%%%%%%%%%%%%%%%%%%%%%%%%%%%%%%%%%%%%%%%%%%%%%%%%%%%%%%%%%%%%%%%%%%%%%%%%%%%%
\begin{figure}[htbp]
\begin{center}
\includegraphics[width=0.45\textwidth]{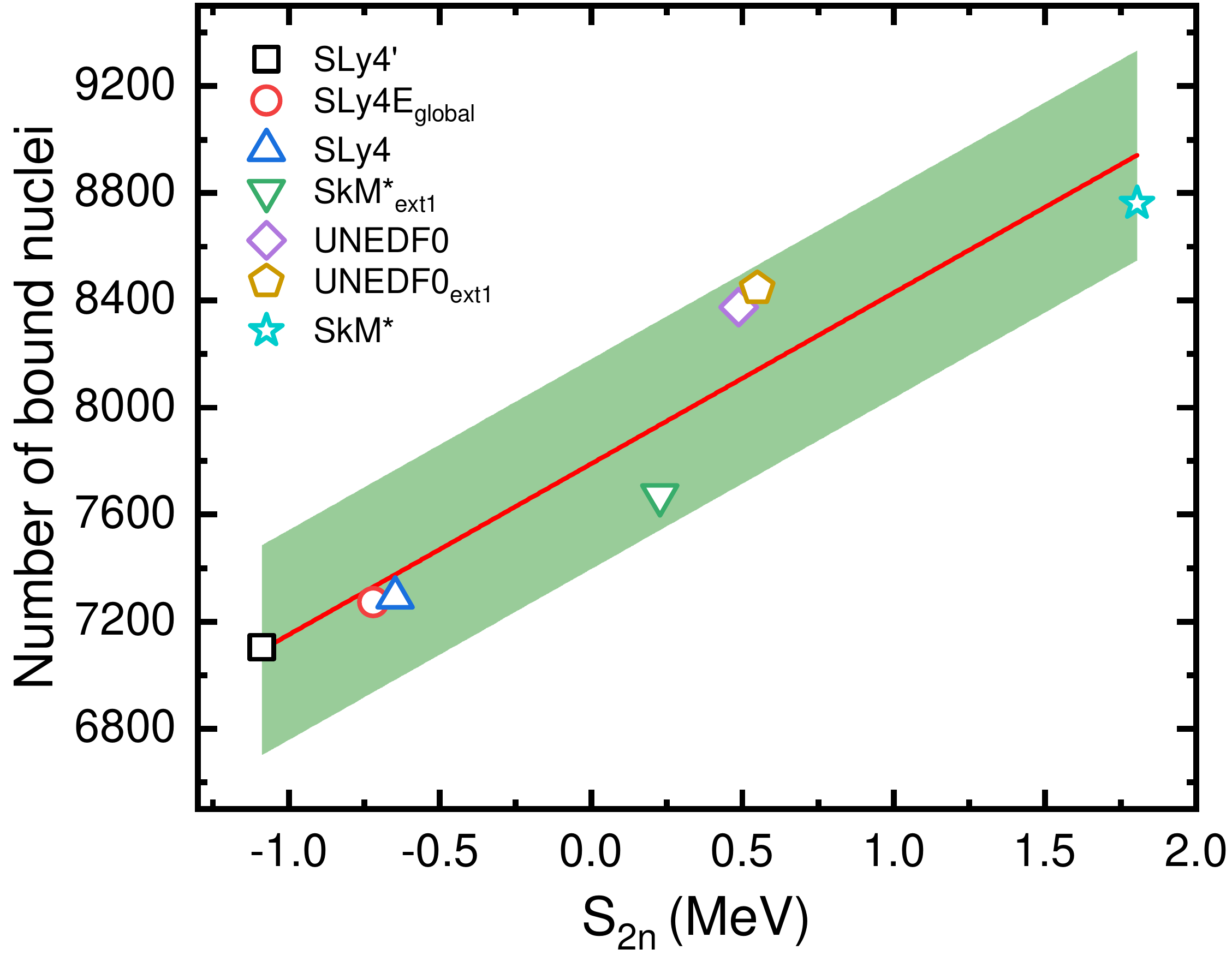}% Here is how to import EPS art
\caption{\label{Fig1}
Calculated $S_{2n}$ of $^{39}$Na with seven Skyrme forces and
the corresponding total number of bound nuclei from $Z$=8 to 120.
The shadows show the confidential interval at 95$\%$.
}
\end{center}
\end{figure}

\begin{figure*}[htbp]
\begin{center}
\includegraphics[width=0.70\textwidth]{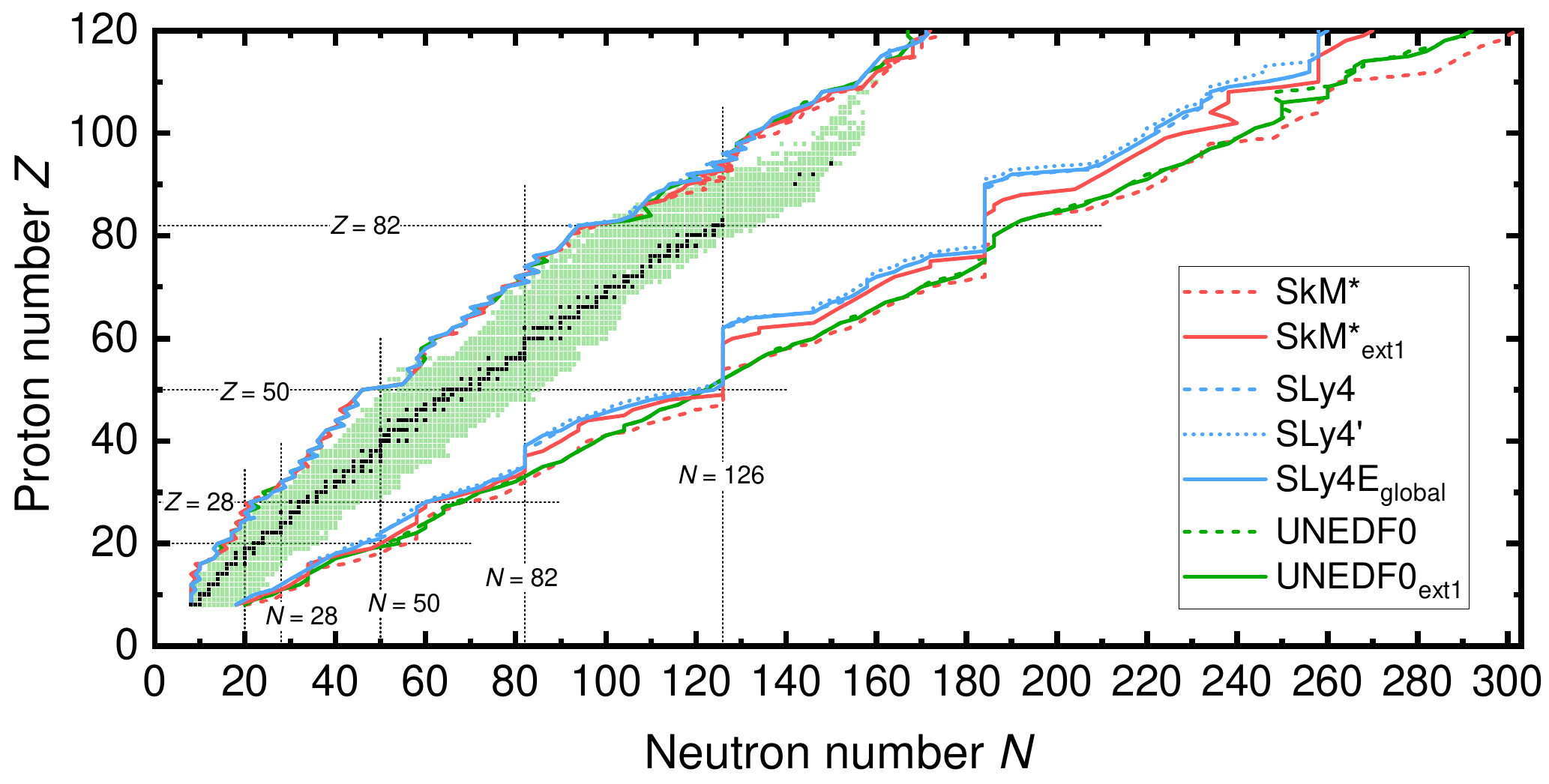}% Here is how to import EPS art
\caption{\label{Fig2}
The calculated nuclear landscape boundaries with seven Skyrme forces (see text).
The black squares denote stable nuclei and green squares denote the experimental observed nuclei
according to ~\cite{ame2020}.
}
\end{center}
\end{figure*}

The recent experiment on $^{39}$Na has attracted great interests for theorists~\cite{Chai2020,Otsuka2020}.
$^{39}$Na has a magic neutron number of $N$=28 but has a well prolate deformation and
a deformed halo structure~\cite{Chai2020}.
The observation of  $^{39}$Na provides a good opportunity for examination of various nuclear mass models.
In Fig.\ref{Fig1}, the two-neutron separation energies $S_{2n}$ of $^{39}$Na
are calculated with seven Skyrme-type forces.
One can see SkM$^*$, SkM$^*$$_{\rm ext1}$, UNEDF0 and UNEDF0$_{\rm ext1}$ forces could reproduce
the existence of $^{39}$Na, while three SLy4-class forces obtain negative $S_{2n}$.
SkM$^*$ gives the largest $S_{2n}$ of  $^{39}$Na and predicts that $^{41}$Na is the drip line nucleus.
Correspondingly, we performed global calculations of nuclear binding energies from $Z$=8 to $Z$=120
with seven Skyrme forces.
The total number of bound nuclei of the nuclear landscape from $Z$=8 to $Z$=120 ranges from 7105 to 8761
with different Skyrme forces.
Generally, we see that the Skyrme force obtains a large $S_{2n}$ of $^{39}$Na  also predicts a large number of
bound nuclei.
The $S_{2n}$ of $^{39}$Na is strongly correlated with the total number of bound nuclei $N_b$, with
a correlation $r$=0.947. The linear regression gives $N_{b}\thicksim\mathcal{N}$(b + a${S_{2n}}$, $\sigma^2$),
in which $a$=638.5, $b$=7789.7, and $\sigma$=199.9.
Once we know the experimental $S_{2n}$ of $^{39}$Na, we can immediately get a stringent prediction of the total
number of bound nuclei of the nuclear landscape according to this linear regression.

Fig.\ref{Fig2} displays the calculated nuclear landscape boundaries with seven Skyrme forces.
Large uncertainties in nuclear landscape boundaries are shown in neutron drip lines.
Furthermore, it can be clearly seen that uncertainties of boundaries in light and medium mass region are small
but propagate to remarkable uncertainties in boundaries of heavy and superheavy mass region.
The consistency between the  $S_{2n}$ of $^{39}$Na and landscape boundaries is shown.
SkM$^*$ results in the furthest extension of neutron drip line, while
SLy4 results in the nearest boundaries. UNEDF0 results are close to that of SkM$^*$.
SkM$^*$$_{\rm ext1}$  boundaries are between SLy4 and SkM$^*$, UNEDF0 results.
The recent Bayesian mixing of eleven mass models infers that the total number of bound nuclei is
7708$\pm$534~\cite{Bayesian2020}.
This Bayesian-mixing inference is very close to the SkM$^*$$_{\rm ext1}$ prediction of 7671 as constrained by
newly observed $^{39}$Na.
Present calculations employ the HO basis while coordinate space calculations should be more accurate but are too costly.
For example, with SkM$^*$$_{\rm ext1}$,   $S_{2n}$ of $^{39}$Na by calculations in HO basis~\cite{Perez2017}
and in coordinate space~\cite{hfbax2008} are 0.23 MeV and 0.27 MeV, respectively.

It should be noted that SkM$^*$ systematically overestimates binding energies of
neutron-rich nuclei~\cite{Erler2012,zzw2018}.
Thus SkM$^*$ is expected to overestimate the extension of neutron drip line
and its prediction can be seen as an upper limit of nuclear landscape boundaries.
In the literature, similar conclusions can be obtained that SkM$^*$
gives the largest number of bound nuclei while SLy4 gives the smallest number of bound nuclei~\cite{Bayesian2020}.
The symmetry energy a$_{sym}$ at the saturation density $\rho_0$ may paly a role.
However, the extended SkM$^*$$_{\rm ext1}$ has a close a$_{sym}$ to that of SkM$^*$.
It has been pointed out that the symmetry energy at $\frac{2}{3}\rho_0$ (0.11 fm$^{-3}$) is strongly correlated with
the neutron drip line location~\cite{ChenLW2015}.
Indeed, the symmetry energies at subsaturation (0.11 fm$^{-3}$) are 26.90, 26.54, 26.49, 25.69, 24.70, 24.37, 24.31 MeV
for SLy4$'$, SLy4$_{\rm global}$, SLy4, SkM$^{*}$$_{\rm ext1}$, UNEDF0, UNEDF0$_{\rm ext1}$, SkM$^{*}$, respectively.
These are strongly  correlated with the total number of bound nuclei $N_b$, with
a correlation $r$=$-$0.989.
This exactly verified that the total number of bound nuclei is correlated with symmetry energy at subsaturation.
We pointed out that SkM$^*$$_{\rm ext1}$ is a very reasonable force to describe the drip line nuclei
around $^{39}$Na and the nuclear landscape boundaries.

The associated $r$-process paths vary with different models, which is defined
as $S_{2n}\approx$2.0 MeV~\cite{Erler2012,ChenLW2015}.
The kink patterns of $r$-process paths and boundary lines appear around neutron magic shells.
Generally, the boundary lines have strong  odd-even effects.
For each isotope, the number of bound nuclei $N_{drip}$  can be determined as a function of charge number $Z$.
In Fig.\ref{Fig2}, for different Skyrme forces,  the uncertainties in $N_{drip}(Z)$ are particularly
large just after the neutron magic number while become much reduced towards the next neutron magic number.
This feature is expected to impact the $r$-process uncertainties.

\begin{figure}[htb]
\begin{center}
\includegraphics[width=0.48\textwidth]{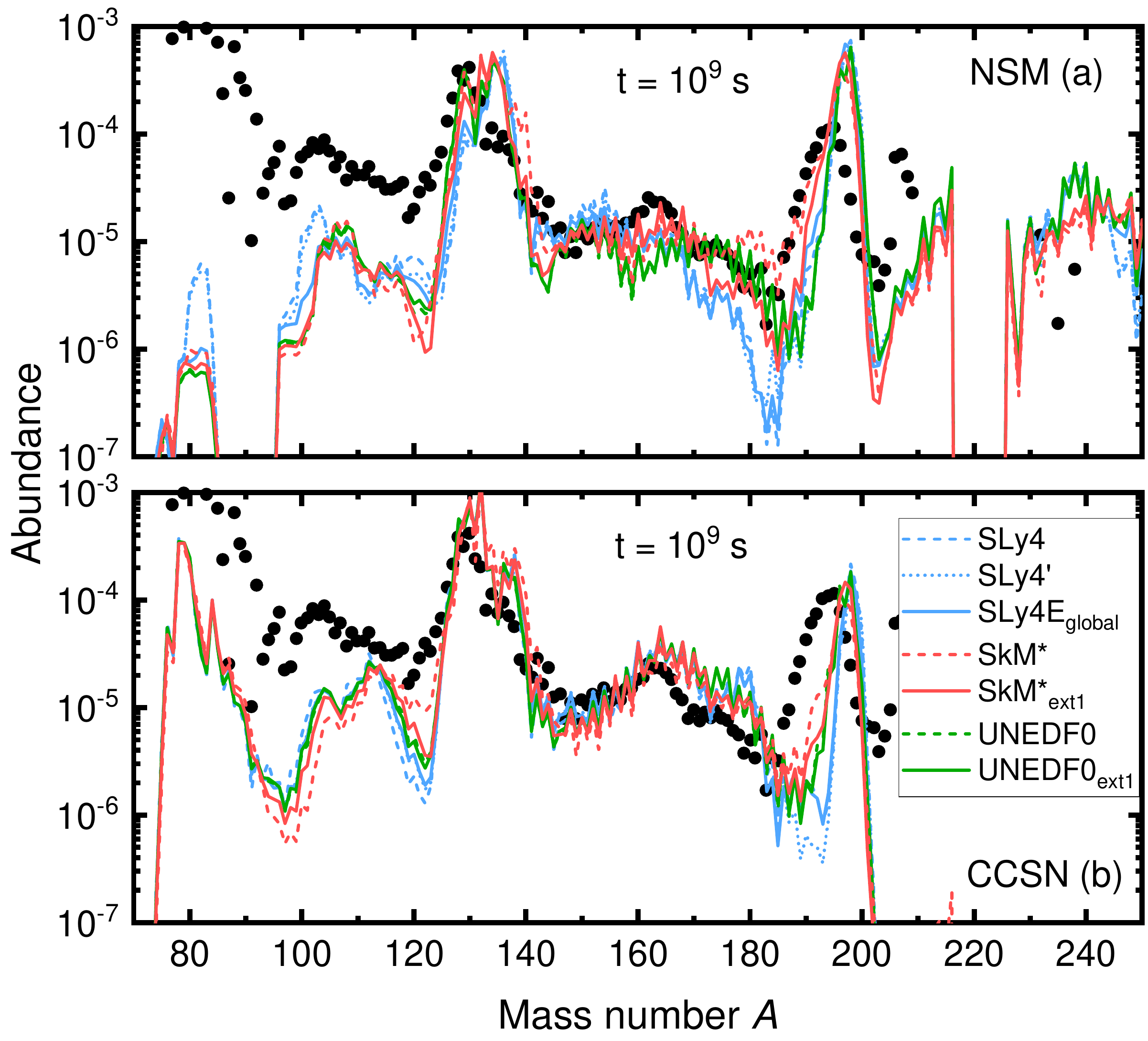}%\\% Here is how to import EPS art
\caption{\label{Fig3}
Calculated $r$-process final abundances as a function of
nuclear mass $A$ with seven Skyrme forces (see text for calculation details).
The $r$-process abundances of NSM scenario (a) and CCSN scenario (b) are displayed.
The solar $r$-process abundance in black dots is also shown for comparison.
}
\end{center}
\end{figure}

The calculated final $r$-process abundances from ejecta of NSM and ejecta of CCSN are displayed in Fig.\ref{Fig3},
based on seven Skyrme forces.
It was known that the prominent abundance peaks around $A$$\thicksim$130 and $A$$\thicksim$195 are
related to the neutron shells at $N$=82 and 126 respectively~\cite{B2FH1957}.
Generally, the resulted uncertainties in NSM scenario are considerably larger than that from CCSN.
In Fig.\ref{Fig3}(a), the most significant uncertainties appear around $A$$\sim$184.
The series of SLy4 forces result in a deep valley in the abundance.
On the other hand, SkM$^*$ produce highest abundance around $A$$\sim$184.
It was explained that this valley is related to nuclear shape transitions in SLy4 calculations~\cite{Martin2016}.
In our results, it seems that the valley varies systematically corresponding to nuclear boundary extensions of nuclear forces.
In other regions, SLy4 gives slightly larger abundances than others around $A$$\sim$140
but smaller abundances around $A$$\sim$130.
In CCSN scenario, the abundances are larger than that of NSM for $A$$<$120,
but much smaller in the region $A$$>$195.
This is reasonable that high neutron density scenario is required to produce the heavy and
superheavy elements.
In both NSM and CCSN cases, the position of the peak around $A$=195 is not well reproduced but
shifted to slightly heavier masses.
In addition, the peaks around  $A$=195 are all overestimated in NSM.
Similar features of the third peak have also been shown in other $r$-process
simulations~\cite{Martin2016,Eichler2015}.

\begin{figure}[htb]
\begin{center}
\includegraphics[width=0.48\textwidth]{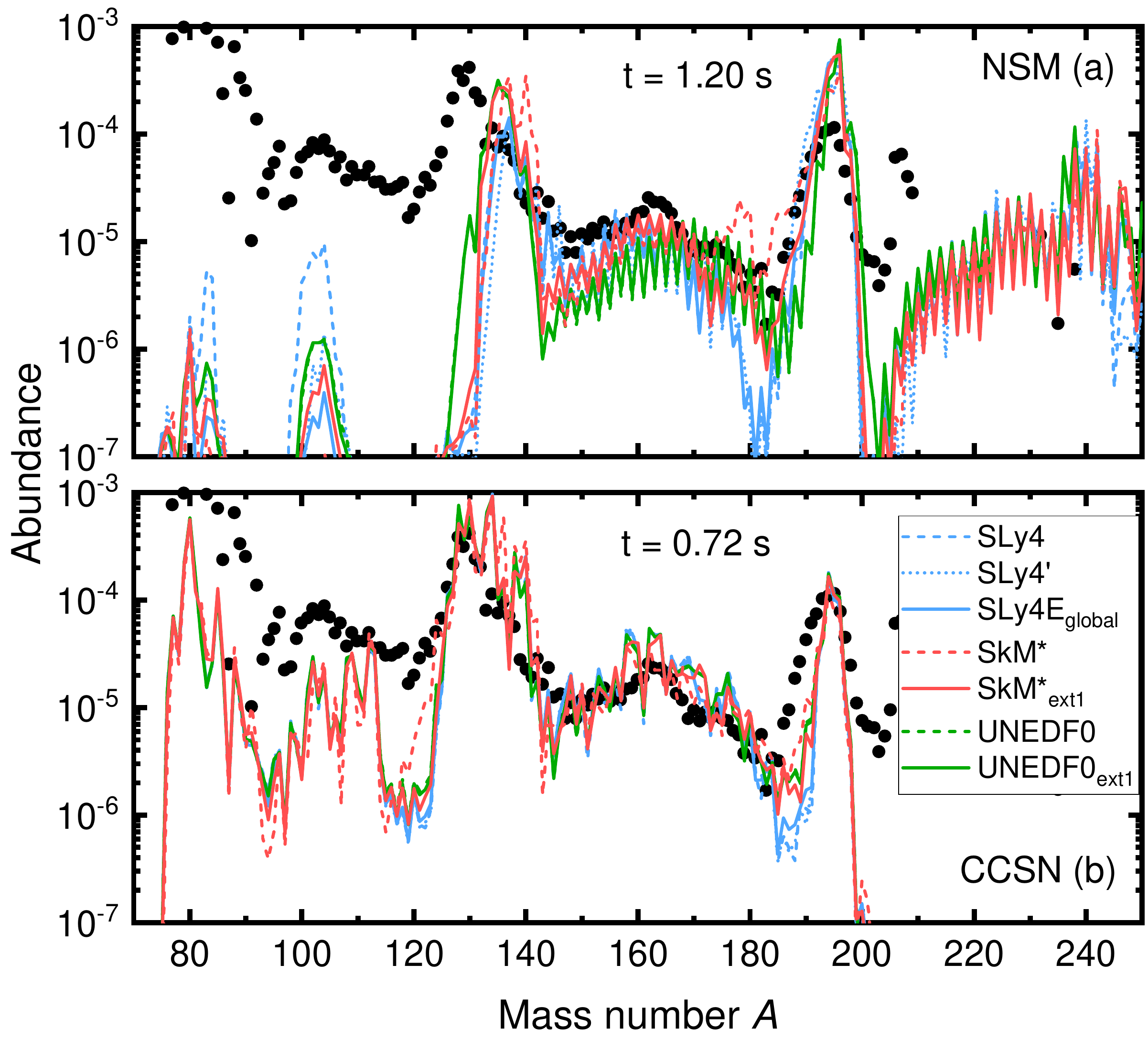}%\\% Here is how to import EPS art
\caption{\label{Fig4}
Similar to Fig.\ref{Fig3} but for $r$-process abundance before freeze-out.
For NSM (a) and CCSN (b), the abundances are obtained at 1.2 s and 0.72 s, respectively.
}
\end{center}
\end{figure}

For detailed analysis of $r$-process evolutions, the abundances during freeze-out are also displayed in Fig.\ref{Fig4}.
The abundances at 1.2 s of NSM and abundances at 0.72 s of CCSN are shown.
In NSM abundance, the significant uncertainties around $A$$\sim$182 present in the early phase, indicating that
the dominate cause is from neutron capture rates close to neutron drip lines.
It can be seen that the position of third peak in NSM is reproduced at 1.2 s but shifted slightly to heavier masses in Fig.\ref{Fig3}.
Indeed, it has been pointed out that the late neutron captures have a direct effect on the
final position of the third peak, with neutrons released from fission of heavier nuclei~\cite{Eichler2015}.
In NSM, the first peak is not yet produced at freeze-out and late fission fragments are essential to
reproduce the first peak around $A$$\sim$130.
Note that both  $N$=82 neutron shell and $Z$=50 proton shell play a role in the first peak.
The evolution analysis indicates that $Z$=50 shell is responsible for the overestimated abundances around $A$$\sim$135 in NSM.
The role of $Z$=50 proton shell is less significant in CCSN since less heavier neutron-rich nuclei
beyond $^{132}$Sn contributed.
The CCSN can reach freeze-out more quickly with less neutron seeds and the role of fission is not significant.
Generally the freeze-out abundances are much irregular and have strong odd-even effects in contrast to final abundances.
The $\beta$ decays and $\beta$-delayed decays in late phases would smooth out the abundances.

%Compared to NSM condition, another site as magnetically driven jets from
%collapsars has also illustrated in Fig.~\ref{Fig4rpccsn}.

\begin{figure}[htb]
\begin{center}
\includegraphics[width=0.48\textwidth]{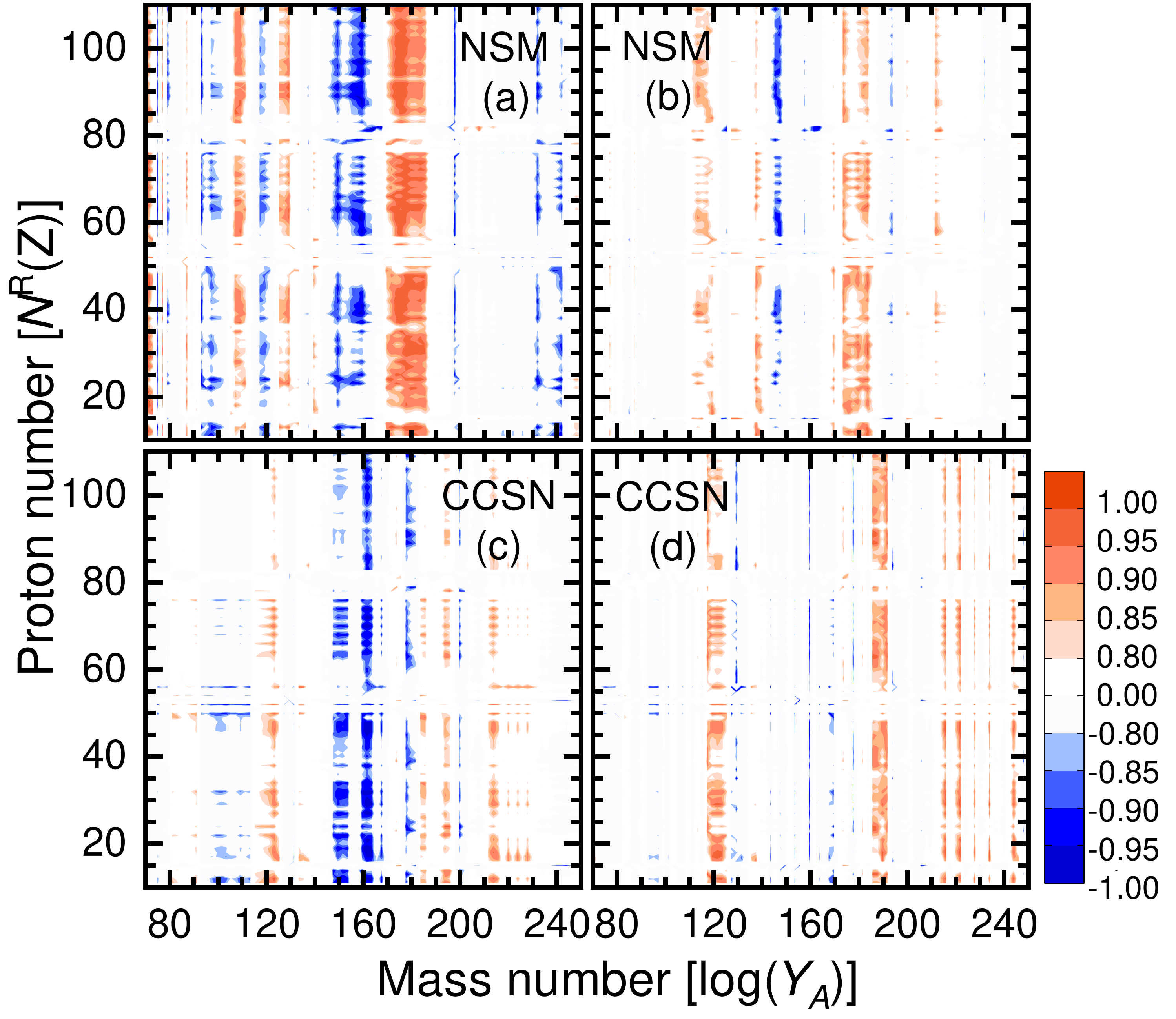}% Here is how to import EPS art
\caption{\label{Fig5}
The calculated correlation matrix between the $r$-process abundance and the number of bound isotopes.
The results for NSM final abundance (a), NSM freeze-out abundance (b), CCSN final abundance (c), CCSN freeze-out abundance (d)
are shown.
In the correlation matrix, the logarithm of abundances $log(Y(A))$ and the relative value of $N_{b}$($Z$)/$Z$ are used, where
$N_{b}$($Z$) is the number of bound isotopes for each $Z$.
}
\end{center}
\end{figure}

Finally, the statistical analysis is performed to look into the correlations
between neutron drip lines and $r$-process abundances.
The covariance correlation matrix is shown in Fig.\ref{Fig5}.
In the correlation analysis, the logarithm of abundances $log(Y(A))$ in terms of nuclear mass $A$ is adopted.
For the other side, the relative value $N^R$($Z$)=$N_{b}$($Z$)/$Z$ is used, where $N_{b}$($Z$)
is the number of bound isotopes
of each charge number $Z$. The relative uncertainties emphasize the correlations between drip-line
light nuclei and $r$-process since drip-line heavy nuclei are not likely accessible.
The correlation matrix is calculated as,
%\begin{eqnarray*}
\begin{flalign}
    &~~{\rm Corr}[{\rm log}Y(i),N^R(j)]&
    \nonumber \\ &=
    \frac{  \frac{1}{6}
    \sum_{k=1}^{7}
    [{\rm log} Y(i,k) - \overline{{\rm log}Y(i)}] \cdot
    [N^R(j,k) - \overline{N^R(j)}]}{
    \sqrt{ \sigma[{\rm log}Y(i)]^2 \sigma[N^R(j)]^2   }}
    &
    \nonumber
                                                                  \label{eqn.05}
\end{flalign}
%\end{eqnarray*}
where $k$ denotes the results of seven Skyrme forces and $\sigma^2$ denotes the variance.
In the NSM case, we found strong positive correlations between the $A$$\sim$180 abundance
and neutron drip lines.
This demonstrated that SLy4 with least extended nuclear  boundaries
would result in the underestimated $r$-process abundance around $A$$\sim$180.
This correlation is not an accident.
The correlation matrix points out that boundaries of some isotopes are especially important.
For example, the drip lines at $Z$=11-13 are important, and the next is $Z$=18.
The analysis of evolution movies (see Supplemental Material) shows distinct features between SkM$^*$ and SLy4 in
the early phase.
The $r$-process with  SkM$^*$ runs very quickly to heavy masses
and considerable abundances are already accumulated just at the left side of the neutron magic number.
It is understandable that SkM$^*$ with most extended boundaries has large early neutron capture rates.
In contrast, SLy4 obtains much less abundances before the third peak due to much less early abundances
just before $N$=126.
The in-between SkM$^*$$_{\rm ext1}$ obtains reasonable abundance in the NSM scenario.
In all cases, it is surprising to see that there is no correlation between $r$-process abundance and neutron drip lines around proton shells at $Z$=50 and 82.
The early abundances of NSM has similar but smaller correlations in Fig.\ref{Fig5}(b) due to larger variances, compared to that of final NSM abundance in Fig.\ref{Fig5}(a).
Fig.\ref{Fig5}(c, d) shows that CCSN cases have no significant correlations between neutron  drip lines and abundances around $A$$\sim$180
in the less neutron-rich environment.
There are some negative correlations in the transitional region around $A$=150$\sim$160.
For some regions, such as the peaks around 130 and 195, there is no strong statistical correlations, since shell effects are dominated.
It is encouraging that the statistical analysis can provide reasonable clues.
In fact, $r$-process evolution is so complex that a big data net analysis is inspiring  from
a different perspective~\cite{Liu2020}.

%%%%%%%%%%%%%%%%%%%%%%%%%%%%%%%%%%%%%%%%%%%%%%%%%%%%%%%%%%%%%%%%%%%%%%%%%%%%%%%%

%%%%%%%%%%%%%%%%%%%%%%%%%%%%%%%%%%%%%%%%%%%%%%%%%%%%%%%%%%%%%%%%%%%%%%%%%%%%%%%%
\section{summary}
%%%%%%%%%%%%%%%%%%%%%%%%%%%%%%%%%%%%%%%%%%%%%%%%%%%%%%%%%%%%%%%%%%%%%%%%%%%%%%%%

In summary, we studied $S_{2n}$ of $^{39}$Na with seven Skyrme forces to constrain
the neutron drip lines.
We found strong linear correlation between $S_{2n}$ of $^{39}$Na and the total number of
bound nuclei.
The in-between SkM$^*$$_{\rm ext1}$ predicts 7671 bound nuclei of the nuclear landscape, which
is very close to the recent Bayesian mixing result.
Our key motivation is to study the uncertainty propagation from neutron drip lines of light nuclei to heavy nuclei,
which is crucial for $r$-process simulations but not accessible by terrestrial experiments.
Based on obtained nuclear masses with different Skyrme forces, $r$-process abundances
from ejecta of NSM and CCSN are calculated using the reaction rate code TALYS and reaction network code SkyNet.
We see large uncertainties in NSM abundances before the third peak.
Further covariance analysis indicate that the abundance uncertainties  are strongly correlated
with the extension of neutron drip lines.
SLy4 predicts the least extended nuclear boundaries and results in the valley in abundances before the third peak.
The statistical analysis shows that neutron drip lines of some isotopes are especially important to constrain
$r$-process in neutron-rich environments.
Our study highlights the further experimental study of $S_{2n}$ of $^{39}$Na would be very needed.
In contrast, the $r$-process in CCSN is not sensitive to the neutron drip lines.
Currently, the understandings of $r$-process  still need comprehensive and accurate nuclear inputs,
in particular, reliable fission predictions.
The statistical analysis can provide reasonable clues and big data analysis  is a promising perspective.
It is reciprocal to develop highly accurate effective nuclear forces for consistent modelings of
Equation of State, drip line nuclei, and nuclear reactions,
for better exportations of nuclear astrophysics at extreme conditions.

%%%%%%%%%%%%%%%%%%%%%%%%%%%%%%%%%%%%%%%%%%%%%%%%%%%%%%%%%%%%%%%%%%%%%%%%%%%%%%%%
\acknowledgments
We are grateful for useful discussions
with J. Lippuner, L.W. Chen,  Y.S. Chen and F.R. Xu.
%are gratefully acknowledged.
This work was supported by the National Natural Science Foundation of China under Grants
No. 12047504, 11975032, 11835001, and 11961141003,
by the China Postdoctoral Science Foundation under Grant No. 2020M670012,
an by the National Key R$\&$D Program of China (Contract No. 2018YFA0404403).
We also acknowledge that all computations in this work were performed in Tianhe-1A
supercomputer located in Tianjin.

\end{document}